\documentclass[12pt]{elsarticle}

\usepackage{amsmath} 	
\usepackage{amsfonts} 	
\usepackage{graphicx} 	
\usepackage{physics}		
\usepackage{gensymb}		
\usepackage{pgfplots}	
\usepackage{subfig}		
\usepackage{bm}
\usepackage{multirow}
\usepackage{float}
\usepackage{colortbl}
\usepackage{color,hyperref}
\usepackage{bookmark}
\usepackage{url}
\usepackage{color,soul}

\journal{Optics Communications}

\pgfplotsset{compat=1.5} 

\begin{document}

\begin{frontmatter}

\title{Measuring correlations in non-separable vector beams using projective measurements}

\author{Keerthan Subramanian}
\address{Centre for Interdisciplinary Sciences \\
Tata Institute of Fundamental Research\\
Hyderabad-500075}

\author{Nirmal K. Viswanathan}
\cortext[mycorrespondingauthor]{Corresponding author}
\ead{nirmalsp@uohyd.ac.in}
\address{School of Physics\\
University of Hyderabad\\
Hyderabad-500046}



\begin{abstract}
Doubts regarding the completeness of quantum mechanics as raised by Einstein, Podolsky and Rosen(EPR) have predominantly been resolved by resorting to a measurement of correlations between entangled photons which clearly demonstrate violation of Bell's inequality. This article is an attempt to reconcile incompatibility of hidden variable theories with reality by demonstrating experimentally a violation of Bell's inequality in locally correlated systems whose two degrees of freedom, the spin and orbital angular momentum, are maximally correlated. To this end we propose and demonstrate a linear, achromatic modified Sagnac interferometer to project orbital angular momentum states which we combine with spin projections to measure correlations.

\end{abstract}

\end{frontmatter}

\section{Introduction}
\label{sec:Intro}

EPR in 1935\cite{einstein1935can} raised concerns about the completeness of quantum mechanics by considering a thought experiment consisting of correlated quantum particles and pointed out how such a system can be used to demonstrate an apparent violation of the uncertainty principle which along with the superposition principle is the bedrock of quantum mechanics. Their experiment lead to two incompatible conclusions that either quantum mechanics was incomplete or that instantaneous non-local correlations which Schr{\"o}dinger called entanglement\cite{schrodinger1935discussion} are a reality. They made use of position and momentum of the particles as observables but its implementation was challenging as these observables span an infinite dimensional Hilbert space. Subsequently Bohm\cite{bohm1951quantum} suggested a more amenable version of the experiment utilizing spin-1/2 particles whose spin states span a 2-dimensional Hilbert space. However these developments had to await Bell\cite{bell1964einstein} who showed that such states seem to have strong correlations which cannot be explained by taking a recourse to hidden variables. Thus the stage was set for Clauser, Horne, Shimony and Holt(CHSH)\cite{clauser1969proposed} and Aspect et.al\cite{aspect1975proposed, aspect1976proposed, aspect1982experimental} to perform experiments to see if such systems show correlations violating Bell's inequality by preparing photons in entangled states and performing projective measurements to obtain correlations. It turned out that reality was indeed bizarre after all and that the work of EPR had unwittingly closed the door on local realism as further attested by recent loophole free tests\cite{hensen2015loophole,giustina2015significant,shalm2015strong}.
With this background to the completeness debate of quantum mechanics, our work looks at experimentally demonstrating a realization of locally correlated degrees of freedom(DoF) of photon states which show strong correlations, culminating in the violation of Bell's inequality. Thus, in addition to reconciling incompatibility of hidden variable theories with reality without recourse to entanglement, we demonstrate that seemingly macroscopic coherent entities like laser beams can also show strong correlations.

Entangled particles are represented by a non-separable composite wavefunction \cite{bertlmann2002quantum, susskind2015quantum, ekert1995entangled}. Two such particles whose polarization DoF are entangled have a composite state vector given by 

\begin{equation}
	\label{eqn:TwoParticleEnt}
	\ket{\psi} = \frac{1}{\sqrt{2}}\bigg[\ket{hh} + \ket{vv} \bigg]
\end{equation}

Where $\ket{h}$, $\ket{v}$ indicate the horizontal, vertical polarization states of the two particles. The polarization DoF spans a two dimensional Hilbert space, and any general polarization state can be represented in the $\ket{h}$, $\ket{v}$ basis as 

\begin{equation}
	\ket{\psi_p} = \frac{1}{\sqrt{\abs{a}^2+\abs{b}^2}}\bigg[ a\ket{h} + b\ket{v} \bigg]
\end{equation}

The state of polarization(SoP) of a light beam, consisting of an ensemble of photons, as represented above can be depicted on the Poincar{\'e} sphere\cite{born1999principles} as shown in Figure \ref{fig:Sphere}. All possible polarization states -- linear, elliptical, circular -- achievable are represented as a point on the Poincar{\'e} sphere by specifying its $(\theta,\phi)$ coordinates, with the $S_1$, $S_2$ and $S_3$ components representing the Stokes parameters\cite{stokes1851composition}. 

Instead of two particles entangled in the polarization DoF, recently it was proposed\cite{chen2010single,borges2010bell} and realized\cite{borges2010bell,karimi2010spin} that an ensemble of photons having two DoF, say spin angular momentum(SAM) and orbital angular momentum(OAM), can also be prepared in a non-separable state. While photonic schemes employing correlations between polarization and propagation direction have been demonstrated\cite{michler2000experiments,gadway2008bell}, a scheme utilizing spin and spatial components was demonstrated in neutrons by \cite{hasegawa2003violation}.
The schemes of \cite{chen2010single,karimi2010spin} still utilize entangled photons and a q-plate for heralding preparation of SAM-OAM correlated states and single-photon counting for detection. In addition \cite{chen2010single,borges2010bell} extensively utilize dove prisms for unitary transformations of the OAM state. While \cite{borges2010bell} attempt to project OAM states using a folded Mach-Zehnder interferometer, they utilize a piezo transducer to scan phase variation between the two beams on $\mu s$ time scales and extract maximal violation of Bell's inequality.
Since photon SAM is related to its polarization\cite{beth1936mechanical} while its OAM is related to the mode of the light beam\cite{allen1992orbital}, vector beams possessing non-uniform SoP across their cross section can be used as realizations of such non-separable single particle states but with two DoF\cite{aiello2015quantum, gabriel2011entangling, mclaren2015measuring}. The polarization degree of freedom of the beam can be easily projected using a polarization beam splitter(PBS) and a waveplate. The OAM DoF, needing a phase sensitive technique, has so far been projected using fork gratings\cite{mair2001entanglement}, spiral phase plates\cite{oemrawsingh2005experimental}, spatial light modulators\cite{yao2006observation,karimi2010spin,mclaren2015measuring,li2015classical} and Mach-Zehnder interferometers\cite{borges2010bell}.

The mathematical isomorphism between entangled systems and locally correlated multi-DoF systems leads us to believe that correlation between the polarization and mode DoF should be identical with the correlation in the entangled case. As an experimental demonstration we prepare photons in SAM-OAM correlated states and propose an achromatic interferometric method consisting of only linear optical components to project the modes onto the $\ket{H}(HG_{10})$,$\ket{V}(HG_{01})$ basis followed by a projective measurement in the polarization DoF to obtain $\ket{hH}$, $\ket{hV}$, $\ket{vH}$, $\ket{vV}$ projections leading to the familiar $\cos(2\theta)$ correlation seen for entangled particles culminating in the violation of the CHSH form\cite{bertlmann2002quantum, clauser1978bell} of Bell's inequality. This leads us to conclude that violation of Bell's inequality heralding the completeness of quantum mechanics can also be demonstrated by using non-separable superposition states possessing no hints of non-local `spooky' effects.

\section{Correlations in entangled systems}

\label{sec:MathEntanglement}

The first experimental implementations \cite{clauser1969proposed, aspect1975proposed, aspect1976proposed, aspect1982experimental} of Bohm's version \cite{bohm1951quantum} of EPR gedanken experiment \cite{einstein1935can} made use of entangled photons which were separated and their polarization measured separately. This led to correlations between the photon polarization states showing a $\cos(2\theta)$ variation where $\theta$ is the relative angle between the projective measurement directions on the two photons.

The entangled state given by Equation~\ref{eqn:TwoParticleEnt} tells us that the photon polarization states are completely correlated. Such correlations are preserved even if the polarization projections for the two photons are performed at the same angle $\theta_1$, since the state vector in the $\ket{l_{\theta_1} l_{\theta_1}},\ket{l_{\theta_1} l_{\theta_1+90}},\ket{l_{\theta_1+90} l_{\theta_1}},\ket{l_{\theta_1+90} l_{\theta_1+90}}$ basis remains non-separable:
\begin{equation}
	\label{eqn:EntStateTheta}
	\ket{\psi} = \frac{1}{\sqrt{2}} \bigg[ \ket{l_{\theta_1}l_{\theta_1}} + \ket{l_{\theta_1+90}l_{\theta_1+90}} \bigg]
\end{equation}

In general correlations arising out of projections for first particle at $\theta_1$ and second particle at $\theta_2$ can be immediately gleaned by representing the state vector in terms of the basis states at angle $\theta_1, \theta_1+90$ for particle-1 and $\theta_2, \theta_2+90$ for particle-2. 
\begin{equation}
	\label{eqn:PsiThetaBasis}
	\begin{aligned}
		\ket{\psi} = \frac{1}{\sqrt{2}} 	\bigg[& \cos(\theta_1-\theta_2)\ket{l_{\theta_1}l_{\theta_2}} \\
										& + \sin(\theta_1-\theta_2)\ket{l_{\theta_1}l_{\theta_2+90}} \\
										& - \sin(\theta_1-\theta_2)\ket{l_{\theta_1+90}l_{\theta_2}} \\
										& + \cos(\theta_1-\theta_2)\ket{l_{\theta_1+90}l_{\theta_2+90}} \bigg]
	\end{aligned}
\end{equation}

Thus the probability of detection in the hh, hv, vh and vv ports in terms of the relative angle $\theta$ are
\begin{equation}
	\begin{aligned}
		P_{hh} & = P_{vv} = \frac{1}{2} \cos^2\theta \\
		P_{hv} & = P_{vh} = \frac{1}{2} \sin^2\theta
	\end{aligned}
\end{equation}
These probabilities lead to a correlation(C) which depends on $\theta$ as,
\begin{equation}
	C(\theta) = P_{hh} + P_{vv} - P_{hv} - P_{vh} = \cos2\theta
\end{equation}

Experimentally the probabilities are obtained from coincidence measurements between the two quantum particles -- ie, how many times one obtains counts in the ports hh, hv, vh, vv using single photon counters. This should however not distract us from the fact that the coincidence measure is in fact a projective measurement where the weights of the basis vectors -- $\ket{hh}$, $\ket{hv}$, $\ket{vh}$ and $\ket{vv}$ -- of the composite state are obtained.

\section{Correlated SAM-OAM photon states}

	The above discussion clearly points out to the fact that correlation between two particles violating Bell's inequality arise naturally out of the non-separability of the state vector describing the system. Is it then surprising that such strong correlations are also obtainable from projective measurements on a system whose two degrees of freedom -- which individually span a 2-dimensional and together span a 4-dimensional Hilbert space -- are locally correlated? In fact, this form of local correlations as opposed to non-local multi-particle correlations in quantum mechanics, has been christened Classical Entanglement \cite{spreeuw1998classical,ghose2014entanglement} though we refrain from using it as it has been hotly debated\cite{karimi2015classical}. In addition, hybrid entanglement\cite{barreiro2005generation} demonstrating both local and non-local correlations has also been demonstrated recently \cite{ma2009experimental,karimi2010spin,fickler2014quantum}. In what follows we take liberty in using the term vector beam, mode-polarization correlated beam to be equivalent to SAM-OAM correlated photons and entangled state to connote multi-particle entanglement. 

\begin{figure}
	\centering
	\includegraphics[width=0.6\columnwidth]{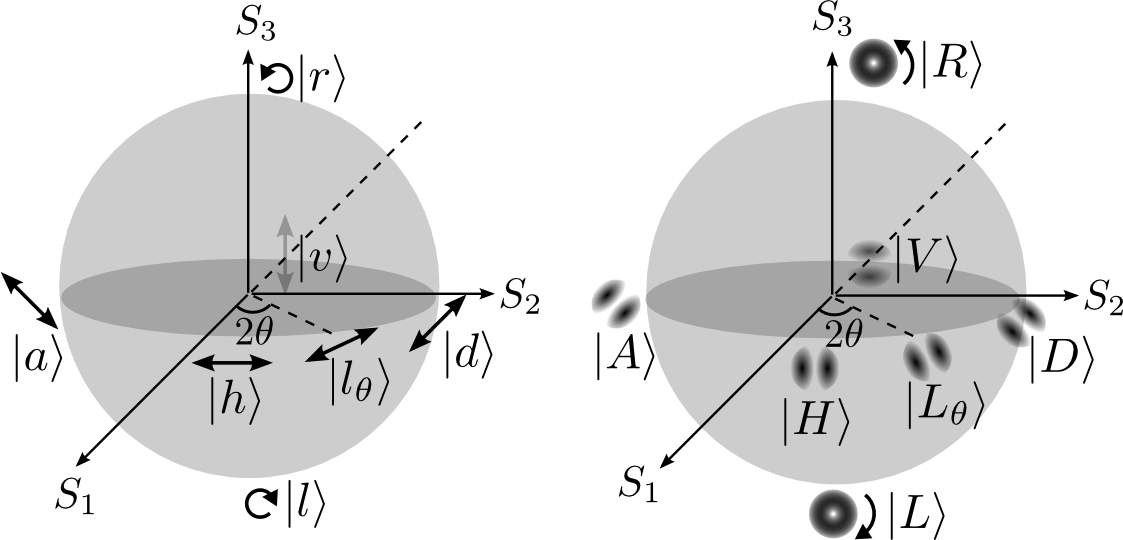}
	\caption{Poincar{\'e} and Padgett spheres for representation of polarization and mode states respectively both of which span a 2-d Hilbert space. Any two diametrically opposite points on the sphere are orthogonal states and can be used to span the entire 2-d Hilbert space}
	\label{fig:Sphere}
\end{figure}

The polarization and mode DoF each span a 2-dimensional Hilbert space and can be represented on a sphere as shown in Figure \ref{fig:Sphere}, with the modal sphere being useful for mapping Hermite Gaussian(HG) and Laguerre Gaussian(LG) modes. In our experiment we use the isomorphism between the two spheres to identify analogous points on the modal sphere\cite{padgett1999poincare} and construct non-separable photon states. Such non-separable states can be realized as vector beams\cite{mclaren2015measuring, d2015hybrid} which have a spatially varying SoP that can be extracted from Stokes polarimetry \cite{goldstein2010polarized, born1999principles}. Since the state of polarization at different points in the beam cross-section is different, a global polarization-mode structure cannot be defined and hence the non-separability ensues. A few state representations of such beams are 

\begin{equation*}
	\begin{aligned}
		\ket{\psi} = \frac{1}{\sqrt{2}} \bigg[\ket{hH} + \ket{vV}\bigg] \\
		\ket{\psi} = \frac{1}{\sqrt{2}} \bigg[\ket{hV} + \ket{vH}\bigg] \\
	\end{aligned}
\end{equation*}

\begin{equation}
	\label{eqn:EntBeam}
	\ket{\psi} = \frac{1}{\sqrt{2}} \bigg[\ket{hR} + \ket{vL}\bigg]
\end{equation}

\section{Projective measurements and correlations}
Let us now consider SAM-OAM correlated photons whose state vector can be described as
\begin{equation}
	\ket{\psi} = \frac{1}{\sqrt{2}} \bigg[\ket{hH} + \ket{vV}\bigg]
	\label{eqn:EntBeamhHvV}
\end{equation}
To measure the correlation between the two DoF, the non-separable vector beam is to be projected simultaneously in the polarization and mode DoF as described below.
\subsection{Polarization/Mode projections}

A polarizing beam splitter(PBS) is a polarization projector whose orientation determines the basis along which an incoming state vector is projected. By itself, without phase shifters, a PBS oriented at an angle $\theta$ can be used to project the state vector along the linearly polarized state $\ket{l_\theta}$. Though functionally the PBS does not operate on the mode DoF, for a vector beam passing through it this does not mean that its output mode is unchanged. 

If we could conceive of an optical device, hereinafter called a mode beam splitter(MBS), described in Section \ref{sec:MBS}, which would analogously operate on the mode DoF and project the modes onto $\ket{H}$,$\ket{V}$ and in general $\ket{\L_{\theta}}$ basis state on the modal sphere we would be able to perform OAM projections.

Combining the polarization and mode projections by passing a beam first through a MBS and then through two PBSes, one at the output of each port as shown in Figure \ref{fig:MBS_PBS}, we can resolve the beam into all its components -- $\ket{hH}$, $\ket{vH}$, $\ket{hV}$, $\ket{vV}$. In general, we can resolve the photon states into $\ket{l_{\theta_1}L_{\theta_2}}$, $\ket{l_{\theta_1}L_{\theta_2+90}}$, $\ket{l_{\theta_1+90}L_{\theta_2}}$, $\ket{l_{\theta_1+90}L_{\theta_2+90}}$ basis states by orienting the MBS along $\theta_2$ and PBSes along $\theta_1$. Such measurements would do away with single photon counting as the ensemble measurements directly give the weights corresponding the the basis states.

\subsection{Correlation between SAM and OAM}

\begin{figure}[h]
	\begin{center}
		\includegraphics[width=0.6\columnwidth]{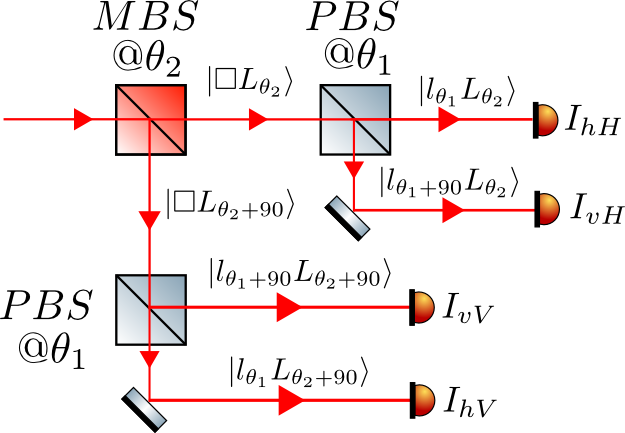} 
	\end{center}
	\caption{Projective measurement scheme for SAM-OAM projection. MBS is a mode beam splitter which projects the OAM state, while PBS projects the polarization or SAM state. The MBS is realized in our experiment as shown in Figure \ref{fig:MBS}}
	\label{fig:MBS_PBS}
\end{figure}

By fixing the orientation of the MBS, and rotating the PBSes one can obtain various projection weights for different relative orientations by measuring the intensity at the four ports as shown in Figure \ref{fig:MBS_PBS}. These measurements give us the analogue of coincidence counts in the Aspect experiment, which lead us to obtain the correlation\cite{bertlmann2002quantum} between the polarization and mode DoF as,

\begin{equation}
	\label{eqn:Correlation}
	C(\theta) = \frac{I_{hH} + I_{vV} - I_{hV} - I_{vH}}{I_{hH} + I_{vV} + I_{hV} + I_{vH}}
\end{equation}

Where the intensities $I_{hH}$, $I_{vH}$, $I_{hV}$ and $I_{vV}$ refer to normalized intensities. Though with MBS and PBS orientations at angles as shown in Figure \ref{fig:MBS_PBS} the state vector is resolved into the $\ket{l_{\theta_1}L_{\theta_2}}$, $\ket{l_{\theta_1}L_{\theta_2+90}}$, $\ket{l_{\theta_1+90}L_{\theta_2}}$, $\ket{l_{\theta_1+90}L_{\theta_2+90}}$ basis states, we still denote the ports as hH, vH, hV and vV for notational convenience.
\vspace{6pt}

By obtaining the correlation coefficient $S(\theta)$ for different relative orientations($\theta$), given in Equation \ref{eqn:CHSH}, one can obtain the CHSH form of Bell's inequality \cite{clauser1978bell} which is obeyed by hidden variable theories.

	\begin{equation}
		\label{eqn:CHSH}
		\begin{aligned}
			-2 \leq S(\theta_1,\theta_2,\theta'_1,\theta'_2) & = C(\theta_1,\theta_2) + C(\theta'_1,\theta_2) + C(\theta'_1,\theta'_2) - C(\theta_1,\theta'_2) \leq +2 \\
			S(\theta) & = 3C(\theta) - C(3\theta) \\
		\end{aligned}
	\end{equation}
\section{Vector beam generation}

\begin{figure*}
	\begin{center}
		\includegraphics[width=0.9\textwidth]{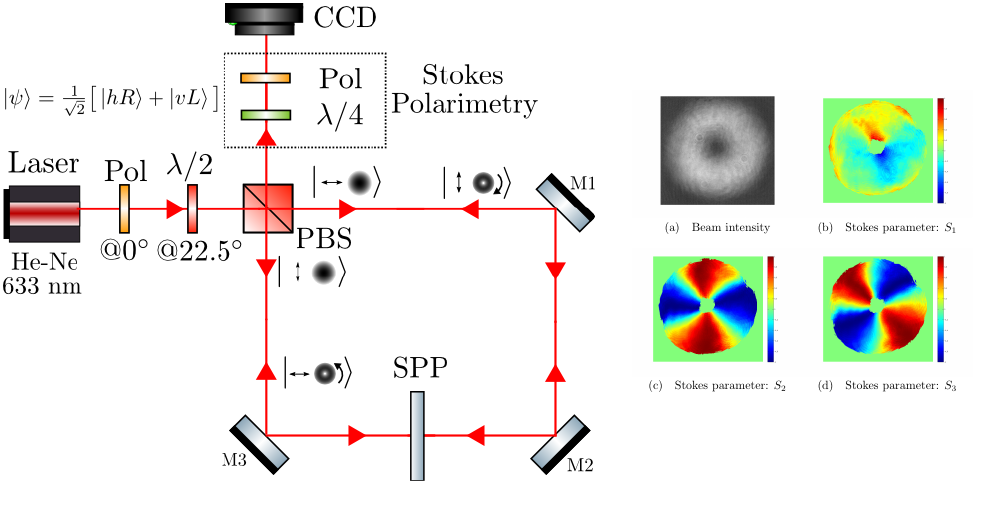}
	\end{center}
	\caption{(Color Online)Vector beam preparation and Stokes parameters. The figure shows the experimental setup used for generating non-separable SAM-OAM states while the inset shows the resulting Stokes tomogram which was obtained experimentally. $M_i$ mirrors; Pol, Polarizer; PBS, Polarizing beam splitter; SPP, spiral phase plate; $\lambda/2, \lambda/4$ - Half and quarter wave plates; CCD, Charge coupled device for imaging}
	\label{fig:EntModeGen}
\end{figure*}

We use a Sagnac interferometer with a spiral phase plate(SPP)\cite{salla2015recovering} as shown in Figure \ref{fig:EntModeGen} to generate a non-separable vector beam. The resulting beam at the output of the interferometer may be described as 

\begin{equation}
	\label{eqn:EntMode}
	\ket{\psi} = \frac{1}{\sqrt{2}} \bigg[\ket{hR}+\ket{vL}\bigg]
\end{equation}

The resulting intensity distribution and polarization variation of the beam, in terms of Stokes parameters($S_1$,$S_2$,$S_3$)\cite{samlan2015quantifying}, obtained experimentally, is shown as  false color images in Figure \ref{fig:EntModeGen}. The beam information so obtained clearly emphasizes the high degree of SAM-OAM correlation.

\section{Mode Beam Splitter: An interferometric implementation}
\label{sec:MBS}
The correlation measurements demand that the input mode of the light beam be projected onto the $\ket{L_\theta}$ basis states. With this requirement in mind and utilizing the fact that any two diametrically opposite points on the modal sphere are orthogonal states and can be used to span the entire 2-dimensional space, we immediately realize that the modes on the equator, the rotated HG modes $\ket{L_{\theta}}$, can be obtained by a linear combination of equal weights of $\ket{R}$, $\ket{L}$ with a suitable phase modulation as shown in Equation \ref{eqn:RLCombination}:
\begin{equation}
	\label{eqn:RLCombination}
	\ket{L_{\theta}} = \frac{1}{\sqrt{2}} \bigg[e^{i\theta}\ket{R} + e^{-i\theta}\ket{L}\bigg]
\end{equation}

\begin{figure*}
	\begin{center}
		\includegraphics[width=0.8\textwidth]{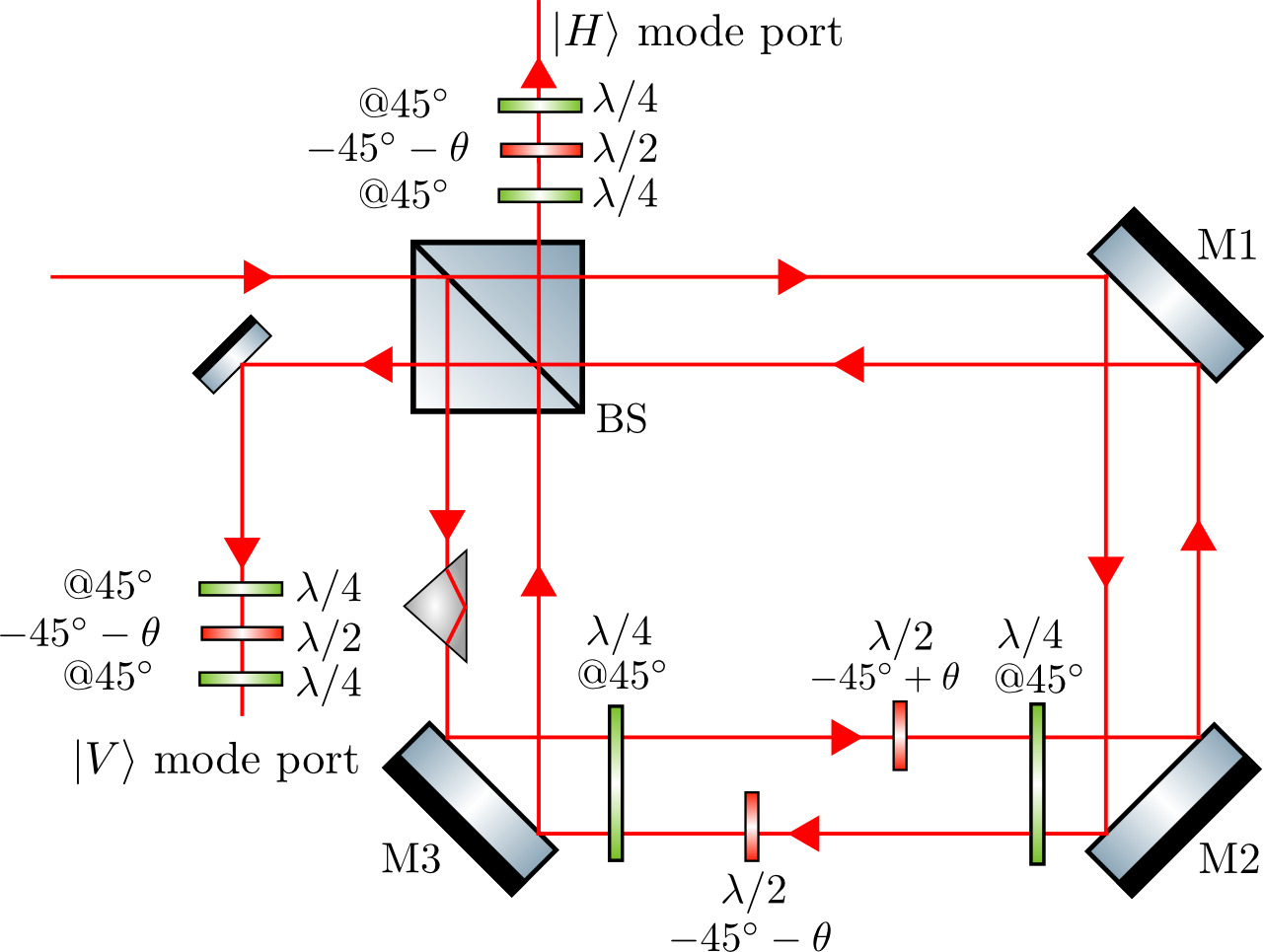}
	\end{center}
	\caption{Achromatic, linear interferometric implementation of an MBS which is used in our experiment to project OAM states; BS, beam splitter; $M_i$, mirrors; $\lambda/2, \lambda/4$ - Half and quarter wave plates. The QHQ plates consisting of $\lambda/4, \lambda/2, \lambda/4$ in that order are used for introducing a geometric phase through transformations on the polarization DoF}
	\label{fig:MBS}
\end{figure*}

Thus in order to resolve the incoming modes into $\ket{H}$ state one needs to superpose the input beam and its twin which has undergone an extra reflection. This ensures that the $\ket{R}$ component of the beam is superposed with its twin, which due to an extra reflection flips its azimuthal phase variation from anticlockwise to clockwise culminating in an $\ket{L}$ mode. Such a scheme can be readily implemented in an interferometer with linear optical components as shown in Figure \ref{fig:MBS}. However, in general to project the mode onto the rotated HG state $\ket{L_{\theta}}$ the beam will need to accrue a phase of $\theta$ while its twin loses the same amount. The phases that the beams accumulate or lose with respect to each other can be dynamical or geometric. Dynamical phase can be manipulated by simply letting the beams travel different path lengths through the same media, while geometric phase can be accumulated by performing \st{unitary} transformations on the polarization state \cite{pancharatnam1956generalized,berry1984quantal,berry1987adiabatic,bhandari1988observation,courtial1999wave}.

The best implementation scheme, with the stability of the interferometer in mind, is to have a common path scheme such as a Sagnac interferometer and introduce opposite geometric phase for the counter-propagating beams. However such a scheme is untenable in a common path configuration as we seek an extra reflection for the twin beam. As a trade-off between stability and manipulating individual beams we propose a shifted version of the Sagnac interferometer\cite{vieira2013spin} as shown in Figure \ref{fig:MBS}. Such a scheme would in addition give us control over phases for individual beams in addition to giving us the extra reflection we seek, which we implement using a prism and could equivalently be implemented using a mirror or a dove prism. Since the two ports of the interferometer are complementary, if one port projects onto the $\ket{L_{\theta}}$ state the other naturally gives the $\ket{L_{\theta+90}}$ projection.

\begin{figure}
	\begin{center}
		\includegraphics[width=0.3\textwidth]{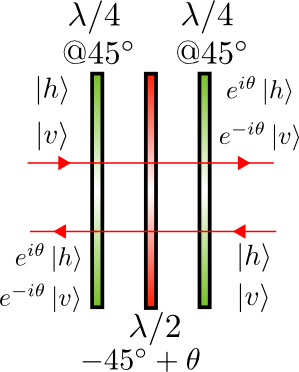}
	\end{center}
	\caption{Action of QHQ for introducing geometric phase through polarization transformations; $\lambda/2, \lambda/4$ - Half and quarter wave plates. These transformations are unitary and restore the polarization to its initial state, thereby ensuring that the polarization DoF of the outgoing beam is identical to the incoming beam}
	\label{fig:GeoPhase}
\end{figure}

In order to implement phase accumulation through geometric phase we utilize a combination of quarter, half, quarter(QHQ) wave plates. When photons with $\ket{h}$, $\ket{v}$ polarization pass through a $\lambda/4$ wave plate oriented at $45\degree$ the SoP is transformed to $\ket{R}$, $\ket{L}$ respectively. On passing this through a $\lambda/2$ wave plate at $-45+\theta$ the SoP is now transformed to $\ket{L}$, $\ket{R}$ via $\ket{L_{\theta}}$ and $\ket{L_{\theta+90}}$ respectively. The final $\lambda/4$ wave plate at $45\degree$ transforms it back to $\ket{h}$, $\ket{v}$. However in this sequence of transformations the magnitude of the solid angle is the same, half of which gives the magnitude of the geometric phase, while the closed path representing the transformations are in an opposite sense -- clockwise in one while anticlockwise in the other beam path. Thus, by having one beam pass through the $\lambda/2$ plate at $-45+\theta$ and the other beam at $-45-\theta$ opposite phases can be accumulated culminating in $\ket{L_{\theta}}$ and $\ket{L_{\theta+90}}$ projections for each of the polarization basis states $\ket{h}$ and $\ket{v}$. The phase accumulation with the QHQ combination for each polarization basis state and direction is summarized in Figure \ref{fig:GeoPhase}. Thus, our MBS implementation utilizes only linear optical elements and is achromatic. We analyse below the resulting transformation for scalar beams and vector beams. Contrary to vector beams which have a non-uniform SoP across their cross section and are realizations of correlated SAM-OAM states, scalar beams have a uniform SoP and are realizations of non-correlated SAM-OAM states.


\subsection{\textbf{Scalar beam $\ket{hR}$}}
For this beam, transformation along the anticlockwise direction accumulates a geometric phase $\theta$ while retaining the mode as $\ket{R}$ while the clockwise beam is transformed into $\ket{L}$ mode and accumulates a phase of $-\theta$. Thus the output beam is

\begin{equation}
	\label{eqn:hR_MBS}
	\begin{split}
		\ket{\psi_o} & = \frac{e^{-i\theta}}{2} \bigg[e^{i\theta}\ket{hR} + e^{-i\theta}\ket{hL} \bigg] \\
					 & =  \frac{e^{-i\theta}}{\sqrt{2}} \ket{hL_{\theta}}
	\end{split}
\end{equation}

\subsection{Vector beam $\frac{1}{\sqrt{2}} \big[ \ket{hR}+\ket{vL} \big]$}
The analysis is the same as above for the $\ket{hR}$ component of the beam. The anticlockwise propagating beam corresponding to the $\ket{vL}$ component accumulates a phase of $-\theta$ while its clockwise twin being transformed to $\ket{vR}$ accumulates a phase of $\theta$. These transformations culminate in an output beam 
\begin{equation}
	\ket{\psi_o} = \frac{1}{2} \bigg[ e^{-i\theta}\ket{h} + e^{i\theta}\ket{v} \bigg] \ket{L_{\theta}}
\end{equation}

\section{Results}

\begin{figure*}[!ht]
\centering
\includegraphics[width=0.73\textwidth]{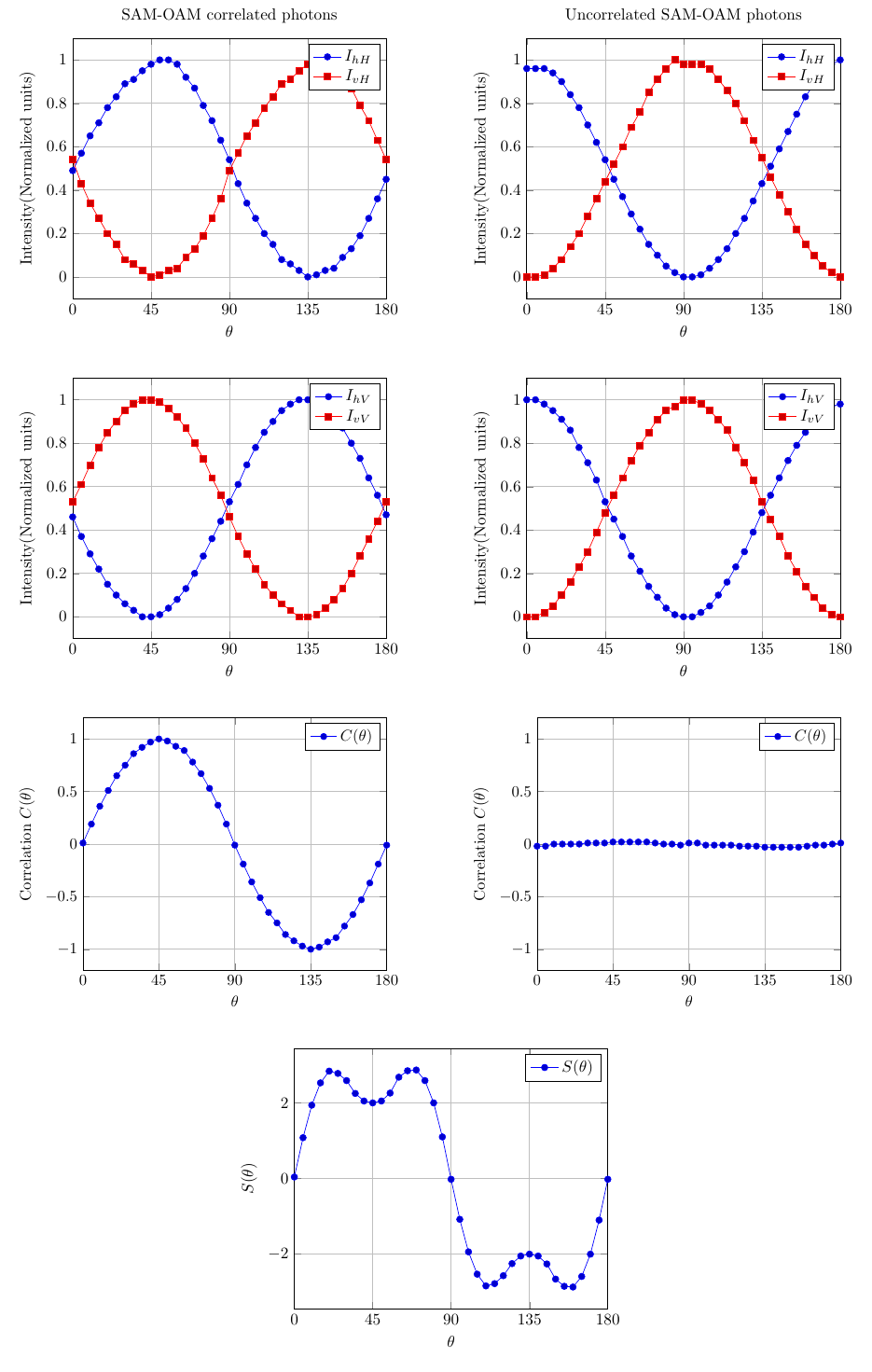}
\caption{Experimental results of projective measurements. Intensity measurements in the four ports of the MBS-PBS setup($I_{hH}, I_{vH}, I_{hV}, I_{vV}$) and correlation[$C(\theta)$] are  contrasted for both the SAM-OAM correlated(column-1) and uncorrelated(column-2) cases. Also presented is the correlation coefficient[$S(\theta)$] calculated using intensity measurements for the correlated case which clearly shows violation of Bell's inequality}
\label{fig:Results}
\end{figure*}

To experimentally perform correlation measurements between the two DoF, a cascade of MBS and PBSes were realized as shown in Figure \ref{fig:MBS_PBS} with the MBS oriented at angle of $0\degree$ and the PBS orientations at $\theta$. Subsequently, beam intensities were measured in all the four ports of the optical setup for both scalar and vector beams. These intensity measurements were put together to obtain the correlation[$C(\theta)$ Equation\ref{eqn:Correlation}] and the correlation coefficient[$S(\theta)$ Equation \ref{eqn:CHSH}]. The experimentally obtained results for scalar and vector beams are contrasted in Figure \ref{fig:Results} from which it can be seen that the intensity and correlation predictions clearly follow the behavior discussed in Section \ref{sec:MathEntanglement}. The correlation measurements for the vector beam show a $\cos(2\theta)$ variation which culminates in the violation of Bell's inequality, while the scalar case is devoid of correlations. It may be pointed out that since the generated correlated state is $\frac{1}{\sqrt{2}} \left[ \ket{hR} + \ket{vL} \right]$ as against $\frac{1}{\sqrt{2}} \left[ \ket{hH} + \ket{vV} \right]$, the kind of non-separable state alluded to in Section(\ref{sec:MathEntanglement}), the $C(\theta)$ and $S(\theta)$ curves in Figure \ref{fig:Results} are shifted by $45\degree$.

\section{Discussion}

What our work highlights is that any two DoF each spanning a 2-dimensional Hilbert space can be made to interact in such a way that they are locally correlated, and that such correlations cannot be described by recourse to hidden variable theories. Since measurements on the two DoF are done locally, we do away with single photon counting and coincidence measurement in order to reckon correlations between them. What is also important to consider is that since we are working with coherent systems, seemingly macroscopic entities such as beams of light appear to violate Bell's inequalities, but such violations are because of the fact that for locally correlated systems that are coherent, projective measurements reduce to average measurements such as measuring the beam power. It is also interesting to point out that introduction of decoherence such as using unpolarized or partially polarized light or mode scrambler in the state preparation destroys the strong correlations and the Bell's inequality is no longer violated. 

A pertinent question in this context, is quantum mechanics non-local? Yes, as unequivocally shown by CHSH and Aspect experiments demonstrating violation of Bell's inequality with entangled photons. However, Bell's theorem answers a more fundamental question of hidden variables which lead to non-locality and not the other way round. So our attempt has been to see if the question of hidden variables can be reconciled using Bell's theorem but without invoking non-locality in any way. If the question of hidden variables is answered, all its consequences however counter-intuitive such as that of non-locality should follow. In conclusion, we hope to have convinced the reader that though non-locality is an oddity and an interesting phenomena in its own right it need not be invoked to demonstrate the completeness of quantum mechanics.

\section*{Acknowledgements}
The authors thank Sir.~M.~V.~Berry for insightful comments on a preliminary version of the manuscript. KS thanks C.~T.~Samlan and D.~N.~Naik for lively discussions and TIFR for fellowship. NKV thanks DST-SERB for continued financial support to this area of research.

\section*{References}
	\bibliography{Bibliography}
\end{document}